\documentclass{article}

\usepackage{PRIMEarxiv}

\usepackage[utf8]{inputenc} 
\usepackage[T1]{fontenc}    
\usepackage{hyperref}       
\usepackage{url}            
\usepackage{booktabs}       
\usepackage{amsfonts}       
\usepackage{nicefrac}       
\usepackage{microtype}      
\usepackage{lipsum}
\usepackage{fancyhdr}       
\usepackage{graphicx}    
\usepackage{orcidlink}
\graphicspath{{media/}}     

\title{Transforming Observations of Ocean Temperature \\ 
with a
Deep Convolutional Residual Regressive Neural Network}

\author{
  Albert Larson \orcidlink{0000-0003-3642-9565} \\
  \texttt{alarson0015@gmail.com} \\
   \And
  Ali Shafqat Akanda \orcidlink{0000-0003-2295-6794}\\
  \texttt{akanda@uri.edu} \\
}

\begin{document}
\maketitle

\begin{abstract}
Sea surface temperature (SST) is an essential climate variable that can be measured via ground truth, remote sensing, or hybrid “model” methodologies. Here, we celebrate SST surveillance progress via the application of a few relevant technological advances from the late 20th and early 21st century. We further develop our existing water cycle observation framework, Flux to Flow (F2F), to fuse AMSR-E and MODIS into a higher resolution product with the goal of capturing gradients and filling cloud gaps that are otherwise unavailable. Our neural network architecture is constrained to a deep convolutional residual regressive neural network. We utilize three snapshots of twelve monthly SST measurements in 2010 as measured by the passive microwave radiometer AMSR-E, the visible and infrared monitoring MODIS instrument, and the in situ Argo dataset ISAS. The performance of the platform and success of this approach is evaluated using the root mean squared error (RMSE) metric. We determine that the 1:1 configuration of input and output data and a large observation region is too challenging for the single compute node and dcrrnn structure as is. When constrained to a single 100 x 100 pixel region and a small training dataset, the algorithm improves from the baseline experiment covering a much larger geography. For next discrete steps, we envision the consideration of a large input range with a very small output range. Furthermore, we see the need to integrate land and sea variables before performing computer vision tasks like those within. Finally, we see parallelization as necessary to overcome the compute obstacles we encountered.
\end{abstract}


\section{Introduction}
Water is both an essential and abundant resource on earth; its availability and quality are critical for sustaining life and ecosystems. Though it is abundant, the majority of earth's water, about 97\%, is found in the ocean, while the remaining 3\% is freshwater found in glaciers, lakes, rivers, and underground aquifers on land. Water is not only crucial for sustaining life, but also plays a vital role in shaping the planet's climate and weather patterns. One significant but understudied climate variable that hydrologists must consider is sea surface temperature. SST has a profound impact on the water cycle, specifically evaporation \cite{oki2006global}. Over-ocean-anomalies like atmospheric rivers can lead to both anomalous and enormous quantities of meteorological water falling on land \cite{dacre2015atmospheric}. The same is true in reverse, as failure of the rains in India are influenced not only by orographic factors from the towering Himalayan mountain range but also determinants of the Bay of Bengal and eponymous nearby ocean \cite{saji2003possible}. Understanding the relationship between sea surface temperature and the water cycle is essential for predicting and adapting to extreme events, managing water resources, and sustainable the global ecosystem \cite{hollmann2013esa}.

Evidence shows that human beings through industrialization have modified and are continuing to significantly modify the climate. However, the modern cause for concern is the rate at which our climate has changed rather than the Boolean of has it or has it not. Measurements of carbon dioxide (CO$_{2}$) tell the story: detected values of atmospheric CO$_{2}$ have increased by 50\% of the starting value at the advent of industrialization \cite{eldering2017orbiting, canadell2007contributions}. Invariant to latitude and longitude, the impacts are felt everywhere. Earth’s response to our stimuli manifests in the form of heat waves, stronger storms, longer periods of drought, greater impulses of meteorological water accumulation over land, and a general increase in environmental variability. While in wealthy communities, modern civil infrastructure serves as a boundary layer to environment-related catastrophes, the poor and powerless are unequally yoked \cite{mekonnen2016four}. One must consider also the importance of the ecology itself. How is the global health of all creatures of the atmosphere, land and oceans \cite{arthington2018brisbane, tharme2003global, dickens2022towards, lynch2023people}? What does the next five, ten, or one hundred years look like at the current rate \cite{sala2000global, vorosmarty2010global}? What changes can be made to mitigate or adapt to implications of past and present poor actions \cite{sanders2020collaborative}? What are the global environmental quality standards \cite{hatanaka2005third}? How can changes be promoted in unequal nation states \cite{keller2021climate,nusrat2022high,messager2022course}? 

SST is worthy of study for a number of reasons: its status as a key observational characteristic of water in the environment; the importance of SST in numerical weather and climate forecasting; SST's detectability via satellite-mounted RS instrumentation; and the availability of matched continuous ground truth temporospatial measurements that can be studied for intercomparison of dataset bias, variances, and uncertainties. We compare the raw satellite observations to the lower resolution but more precise in situ measurements of sea surface temperature (ISAS). We apply a treatment to the lower resolution but generally more available satellite instrument (AMSR-E), setting its target output to be the higher resolution MODIS product. Our hypothesis is that fusing the AMSR-E data to MODIS data will create a product that is closer in performance to MODIS than its AMSR-E input.

SST is a fascinating variable because not only is it a predictor of future weather anomalies, but it represents a largely unexplored stage regarding the capture and storage of CO$_{2}$. Marine carbon dioxide removal (mCDR) is a crucial strategy for mitigating climate change by extracting and sequestering carbon dioxide (CO$_2$) from the atmosphere. Natural processes, such as the biological pump and carbon storage in marine ecosystems, contribute significantly to CO$_2$ removal \cite{honjo2014role, Duarte2013}. Engineered approaches, including iron fertilization to stimulate primary production \cite{Boyd2007}, nutrient optimization for enhanced carbon export \cite{pollard2009southern}, and ocean alkalinity enhancement to increase CO$_2$ absorption capacity \cite{rau1999enhanced}, offer potential avenues for mCDR. However, challenges exist, including environmental risks and unintended ecosystem disruptions \cite{Ciais2013}.

One tool humans have in their arsenal is the ability to artificially model the environment. Modeling is the combination of a priori measurement data with empirically derived functions that simulate and represent the behavior of the environment. Given observed information obtained by sensors or measurement devices, we can use the collective knowledge to make better decisions. The types of observation data we collect about the state of the climate now in a given place, (e.g. tree ring cores, core samples) or via man-made data collection devices (satellites, weather balloons, airplanes, drifters, buoys, gauges), and have better insights to what tomorrow or the same time in five years might bode given current inputs, trajectories, and these empirical functions \cite{piecuch2023river}.

The use of neural networks as a tool for modeling has grown in frequency over the last two decades alongside the increase in availability and speed of fast mathematical computing hardware like graphics processing units. Nvidia has just crossed the trillion dollar market capitalization in large part to the AI boom driven by the proliferation of generative models like GPT and Stable Diffusion \cite{reutersNvidiaBriefly}. Here, we build on an existing neural network architecture called dcrrnn, which stands for a deep convolutional residual regressive neural network and is pronounced "discern" \cite{larson2023discerning, larson2022clearer}. When applied to problems involving the water cycle, dcrrnn is under the umbrella of Flux to Flow (F2F). This is an expectorant measure developed during the creation process of the first experiment. It is hypothesized that one might prefer to extract one or the other depending on the subject environment. Specifically, the extraction of just the dcrrnn structure for use in another vertical. Likewise, development of F2F with a different but related structure and a focus specifically on water-focused variables are anticipated to be better categorized as Flux to Flow projects. Dcrrnn is narrow, whereas F2F is wide. 

The premise and motivation of dcrrnn and F2F is to study measurements of water, be them remotely sensed, gauged measurements, or hybrid datasets. From ingestion, some preprocessing of the data occurs to prepare for training either via statistical conformation or temporospatial constraint. The data is used to train a neural network. Inferences on unseen data are performed to evaluate the trained algorithm according to a relevant figure of merit for the experiment. Our chosen metric in set of experiments is the dimensionless root mean squared error (RMSE). This metric is a solid introductory measurement because of its frequent use and acceptance within the scientific community. Our stochastic framework was successful in mimicking the performance of deterministic algorithms used to predict gauged river streamflow from meteorological forcings. In light of this success and our interest in all facets of the water cycle, we elected to next examine the ability of dcrrnn and F2F to mimic the abilities of image enhancement and ocean modeling software such as the Scatterometer Image Reconstruction (SIR) algorithm, the regional ocean modeling system (ROMS) algorithm, and the Massachusetts Institute of Technology's General Circulation Model (MITgcm) \cite{ROMS,MITgcm,long2015optimum,long1993resolution}. These algorithms take in a variety of different spatial datasets and perform some deterministic process to provide a value-added output. We aimed to evaluate the research questions: "Can dcrrnn, the supervised learning structure, be used as a surrogate to other commonly used image optimization algorithms and ocean modeling software? When applied to SST fields, does dcrrnn improve an input dataset based solely on statistical optimization associated with neural networks and the high resolution training data?"

From the outset, one important facet of dcrrnn and F2F to retain is the integral relationship between the paper and the code used to perform the experiments. A primary motivation behind the work writ large is to disseminate the programming element so others will want to join. Frequently, frustrations and disagreements arise about scientific discourse because the experiment process is not documented, or there is no repeatable proof. We as much as possible take a white box approach. Scripts for these experiments are made available as jupyter notebooks at \url{https://github.com/albertlarson/f2f_sst}. As a primer, we recommend the genesis dcrrnn paper \cite{larson2023discerning}. Though the likely audience is one with or approaching a graduate degree and a penchant for education, our intention is that the language is such that it might be introduced in an undergraduate or advanced K-12 classroom. Coincidingly, pointing out pitfalls is a valuable exercise to help prevent others from repeating the same mistakes. Here, we capture what occurs when a naive, stochastic system is used on a single compute node to attempt a performance of the classic data fusion problem via neural network methods with SST fields as the subject matter. We determine that our approach has definite merit, but requires further investigation with current datasets and other configurations. A prerequisite for advancement of this effort should either be a larger amount of compute time via single node, or via multiple compute nodes conducted in parallel. 

With this article, our contributions to the field are the following: 1, the continued development of F2F as an open source water cycle measurement framework; 2, to further consideration of dcrrnn as a viable neural network architecture; 3, the active constraint of the work to a single compute node and focus on the integration of the manuscript with the code behind the experiments; 4, the consideration and focus on the importance of sea surface temperature from a hydrologist's point of view; 5, a fresh consideration of neural network based data fusion and data engineering techniques; 6, an illustration of the limitations of underparameterization; 7, a demonstration of how the neural network improves its performance when the requirements of the process are relaxed; and last but not least, 8, a biased focus on global water resources.

\section{Materials and Methods}
\subsection{Sea Surface Temperature (SST)}
The origin of SST as a continuously monitored variable began when Benjamin Franklin captured measurements of the ocean as he traversed the Atlantic, acquiring data and synthesizing these observations into information about the Gulf Stream \cite{minnett2019half}. Since then, the field of physical oceanographic research has made great strides in a variety of advancements relevant here such as observational techniques and data analysis methods. In recent decades, satellite remote sensing has emerged as a crucial tool for observing the ocean on a global scale, and for acquiring SST data. Satellites equipped with infrared sensors provide accurate and high-resolution measurements of SST \cite{woolff2007,merchant2012}. These satellite-based measurements offer advantages over traditional in situ measurements, as they provide comprehensive coverage of the oceans, including remote and inaccessible regions. In situ devices are point source measurements and have a relatively limited observation window compared to the hundreds of kilometers sun-synchronous satellites capture at a single moment.

In addition to infrared sensors, satellite sensors that detect microwave radiation, starting with the Nimbus-5 in 1972, have also been used to retrieve SST \cite{gentemann2010passive}. Microwave-based SST retrieval methods have the benefit of low sensitivity to atmospheric conditions, such as cloud cover and aerosols, compared to infrared measurements. The availability of long-term satellite-based SST datasets has led to significant advancements in climate research. Scientists have utilized these fields to study various climate phenomena, including oceanic oscillations such as the Pacific Decadal Oscillation (PDO) and the Atlantic Multidecadal Oscillation (AMO) \cite{knight2006climate,kuo2013decadal}. Large-scale oscillations influence the long-term variability of SST and have implications for regional climate patterns.

Furthermore, satellite-based SST observations have been essential in understanding the impacts of climate change on the oceans. Studies have shown that global warming has led to widespread changes in SST, including increases in average temperatures, alterations in temperature gradients, and shifts in the distribution of warm and cold regions \cite{levitus2005warming}. These changes have significant implications for marine ecosystems, including shifts in species distributions, changes in phenology, and coral reef bleaching events \cite{hoegh2007,kleisner2017marine}. Satellite-derived SST data also contribute to the prediction and forecasting of weather and climate events. The accurate representation of SST conditions is crucial for weather prediction models, as it affects the development and intensity of atmospheric phenomena such as hurricanes and tropical cyclones \cite{monaldo1997satellite,lotliker2014cyclone}. The integration of satellite-based SST data into numerical weather prediction models has improved forecast accuracy, particularly in regions where in situ observations are sparse or nonexistent.

In addition to weather forecasting, satellite-based SST data has practical applications in fisheries management and marine resource monitoring. SST information helps identify optimal fishing grounds by indicating areas with suitable temperature conditions for target species \cite{thorpe2022spatial, hobday2015impacts}. Furthermore, monitoring changes in SST can provide insights into the health of marine ecosystems and aid in the assessment and management of protected areas and biodiversity hot spots \cite{kavanaugh2021satellite,racault2012phytoplankton, chen2020greening}.

\subsection{Aqua}
The Aqua satellite was launched on May 4, 2002. Upon it, two instruments sit: AMSR-E and MODIS. Both, among other things, are designed to study the temperature of the ocean. The measurements obtained as the satellite is moving from South Pole towards North always crosses the equator at approximately 1:30 PM local time nadir (directly below the satellite). In the downward portion of the orbit, the satellite crosses the equator at 1:30 AM local time nadir \cite{parkinson2003aqua}. The AMSR-E instrument ceased functioning after ten years of service. MODIS continues to operate, logging over twenty years of active surveillance. AMSR-E has been succeeded by a follow-on instrument AMSR2. AMSR2 is aboard a Japanese mission called GCOM-W, one of a series of global climate observation missions \cite{gentemann2015situ}. AMSR2 has a slightly larger antenna than AMSR-E, but is similar in scope to the AMSR-E instrument. The de facto replacement of MODIS is VIIRS, an instrument series carried on the Suomi National Polar-orbiting Partnership (SNPP), NOAA-20, and NOAA-21 satellites \cite{brasnett2016assimilating, gladkova2016improved}. Other global aerospace mission datasets associated with SST are available, like the Chinese Haiyang and Gaofen series \cite{liu2016evaluation, meng2019estimating}.

\subsection{AMSR-E}
AMSR-E was a passive microwave radiometer \cite{kawanishi2003advanced, chelton2005global}. The acronym stands for Advanced Microwave Scanning Radiometer for Earth Observing System. There are several products produced on top of the raw radiance data that were collected by this instrument, and the AMSR-E data was processed by different ground stations depending on the parameter of interest. The produced datasets contain latitude, longitude, several physical parameters (e.g., SST, Albedo, soil moisture) as well as other pertinent metadata. As it pertains to sea surface temperature, AMSR-E is available in Level 3 products and as part of Level 4 assimilation system output. As of the writing of this document, Level 2 SST is no longer publicly available. However, we were able to obtain Level 2 fields from the Physical Oceanography Distributed Active Archive Center (PODAAC) through the Jet Propulsion Laboratory before the sunsetting of the data product.

\begin{figure}[!ht]
	\centering
 	\includegraphics[width=1.0\linewidth]{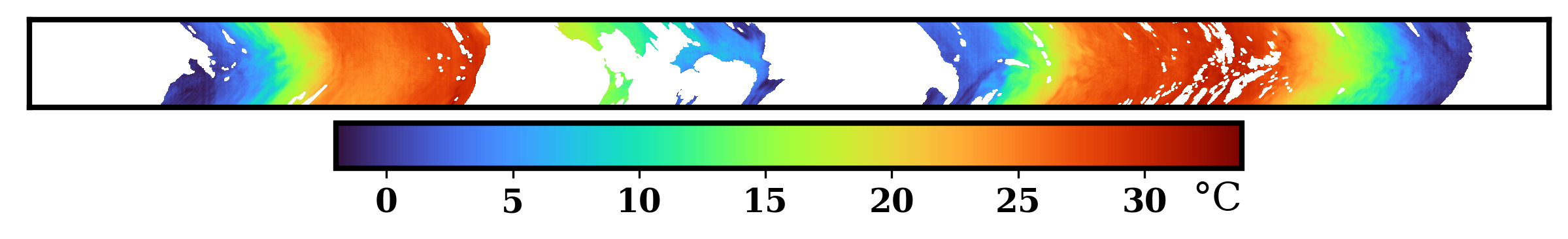}
    \caption{L2 AMSR-E SST field, March 3, 2004, no coordinate system}
    \label{fig2_1}
\end{figure}

To detail a sample, one single Level 2 netCDF (.nc) file containing AMSR-E data was procured. The record selected is that of March 3rd, 2004, with a UTC time of 01:07:01. The file contains three coordinates (latitude, longitude, and time) and thirteen data variables. Each variable is a single matrix comprised of columns and rows of measurements. The important distinction here is that the data structure is stored to reflect the path of the orbit. See Figure \ref{fig2_1}. When the sea surface temperature is plotted as it sits in the matrix, it is difficult to discern what is transpiring. There appears to be some curvature of the measurements, but other than that little is known to an untrained eye beyond the title and colormap.

\begin{figure}[!ht]
	\centering
 	\includegraphics[width=1.0\linewidth]{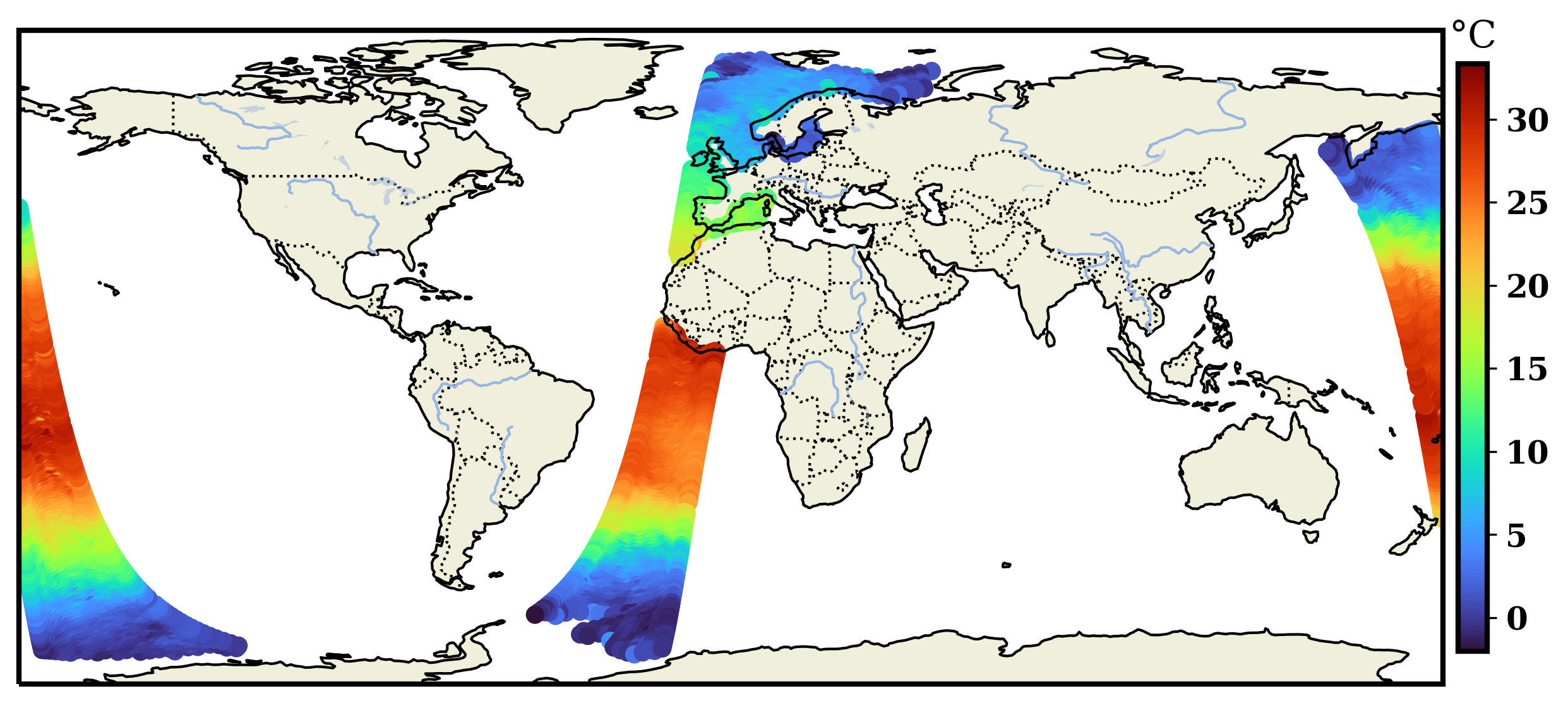}
    \caption{L2 AMSR-E SST, plotted with available coordinates and world map}
    \label{fig2_2}
\end{figure}

Inclusion of the latitude and longitude coordinates, as well as a global basemap generates a clearer picture as seen in Figure \ref{fig2_2}. A single Level 2 AMSR-E SST file contains matrices representing one full orbit around the globe. Each file holds partially daytime and partially nighttime observations. Because of diurnal warming, it is desirable to separate the nighttime and daytime passes. Furthermore, many analyses are comprised of an ensemble of satellite observations from different platforms such as this one. A grid makes for more orderly computations at large spatial scales. Certainly, one could elect to grid every observation to the AMSR-E or MODIS native product coordinate system. With our experiments, we choose the path of rectangular gridding. We consider the Level 3 product because of the interest in spatial relationship across large geographic scales and variable time (daily, weekly, seasonally, yearly, generationally). 

\begin{figure}[!ht]
	\centering
 	\includegraphics[width=1.0\linewidth]{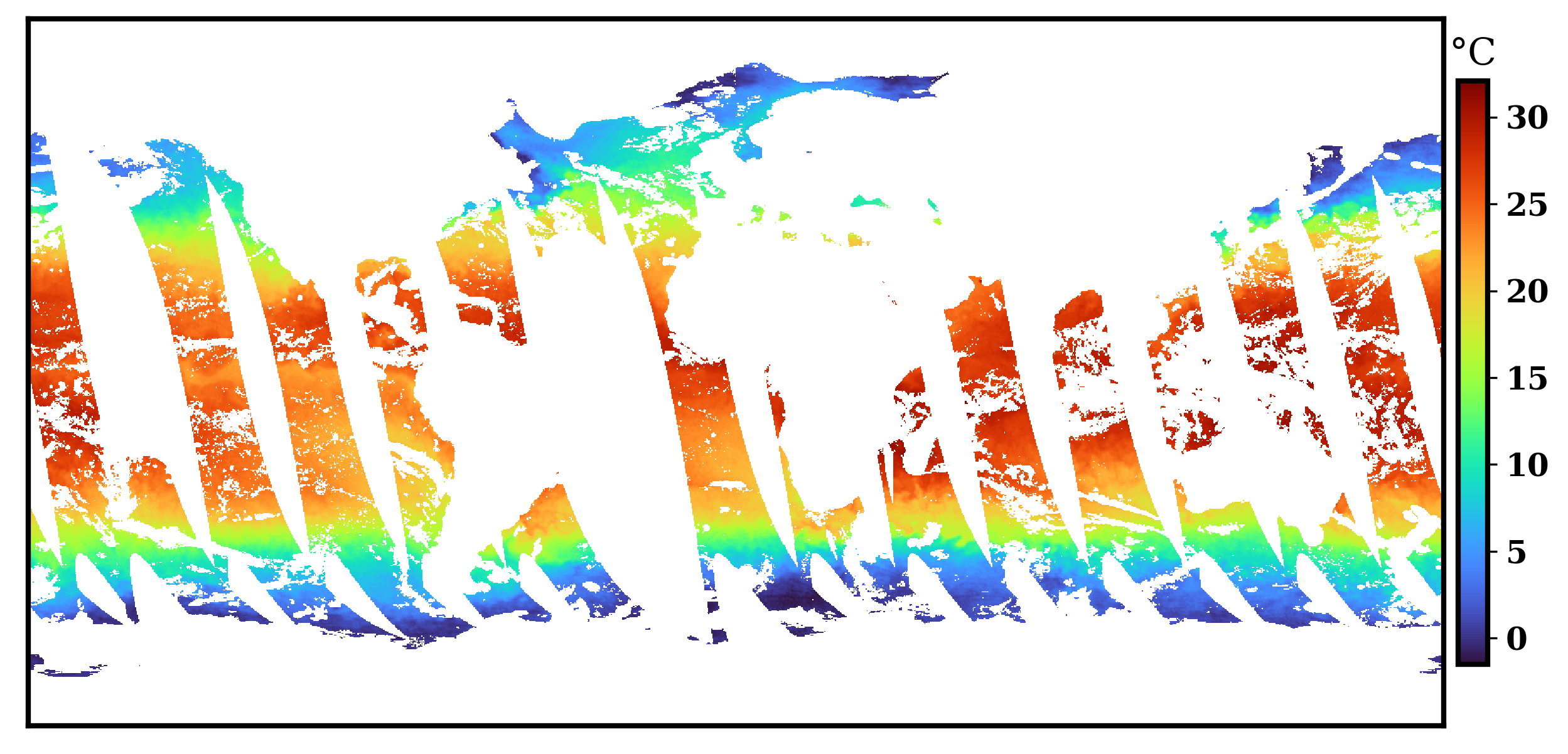}
    \caption{L3 AMSR-E file plotted without supplied coordinate system}
    \label{fig2_3}
\end{figure}

\begin{figure}[!ht]
	\centering
 	\includegraphics[width=1.0\linewidth]{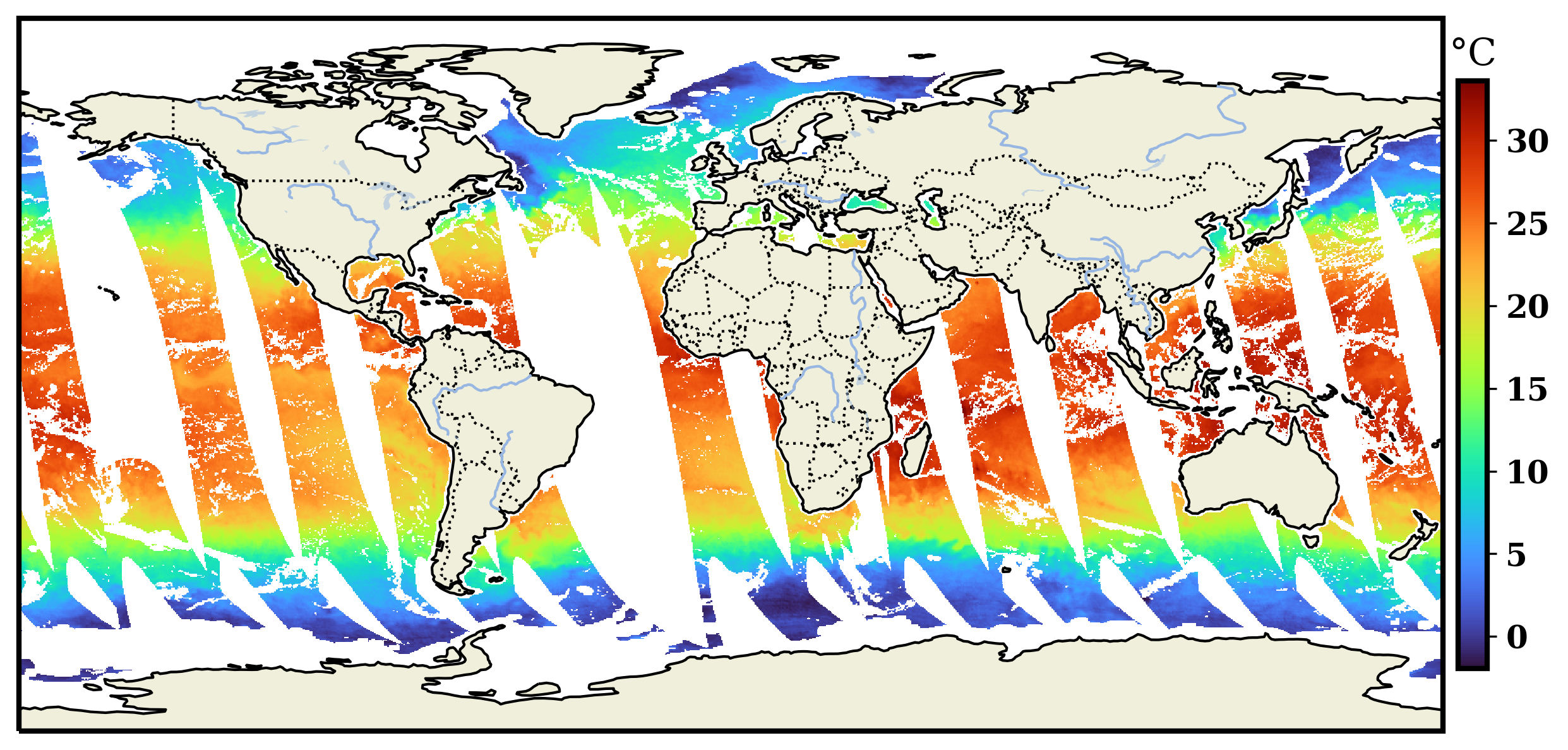}
    \caption{L3 AMSR-E file plotted with available coordinates and world map}
    \label{fig2_4}
\end{figure}

AMSR-E is available (accessed 7 June 2023) via its producer,  Remote Sensing Systems of Santa Rosa, CA \cite{AMSR-E}.  This daily product comes in 25 km resolution and is delineated by daytime and nighttime passes of the satellite. The time series runs from June of 2002 until October of 2011 when the AMSR-E instrument ceased functioning. Figure \ref{fig2_3} illustrates the point that even without explicitly defining the coordinate system in the visualization, the matrix of SST values is already placed in proper spatial order. Figure \ref{fig2_4} reinforces the fact that little change occurs with the inclusion of latitude and longitude coordinates when plotted on a rectangular grid. Here, we simply mean each month's worth of daily daytime and nighttime passes on a pixel-wise basis. We call these day and night in the experiments that follow. We also create a hybrid dataset, where we average the monthly averages of day and night passes together. Finally, we transform all three of these datasets from the native AMSR-E grid system to the slightly different MODIS grid; this function is carried out using the xESMF software \cite{zhuang2019xesmf}.

\subsection{MODIS}
MODIS, or Moderate Resolution Imaging Spectroradiometer, measures thirty-six different radiance bands in the infrared and visible ranges of the electromagnetic spectrum \cite{esaias1998overview}. Level 3 sea surface skin temperature as obtained from MODIS comes in 4 kilometer and 9 km products, and is derived from a subset of the thirty-six radiance bands. The products are available in daily, average of eight days, and monthly products. They are also delineated by daytime and nighttime passes of the Aqua’s polar-orbiting nature. SST products deriving from MODIS are further specified by the length of the waves within the thermal infrared range used to derive the measurement: longer waves (11–12 microns) and middling waves (3–4 microns). The MODIS documentation state that the 3–4 micron wave SST product is less uncertain, but only usable at night because of the daytime sun glint impact on 3–4 micron waves. We use the long wave 11–12 micron infrared measurements to keep constant the source of both daytime and nighttime passes.

\begin{figure}[!ht]
	\centering
	\includegraphics[width=1.0\linewidth]{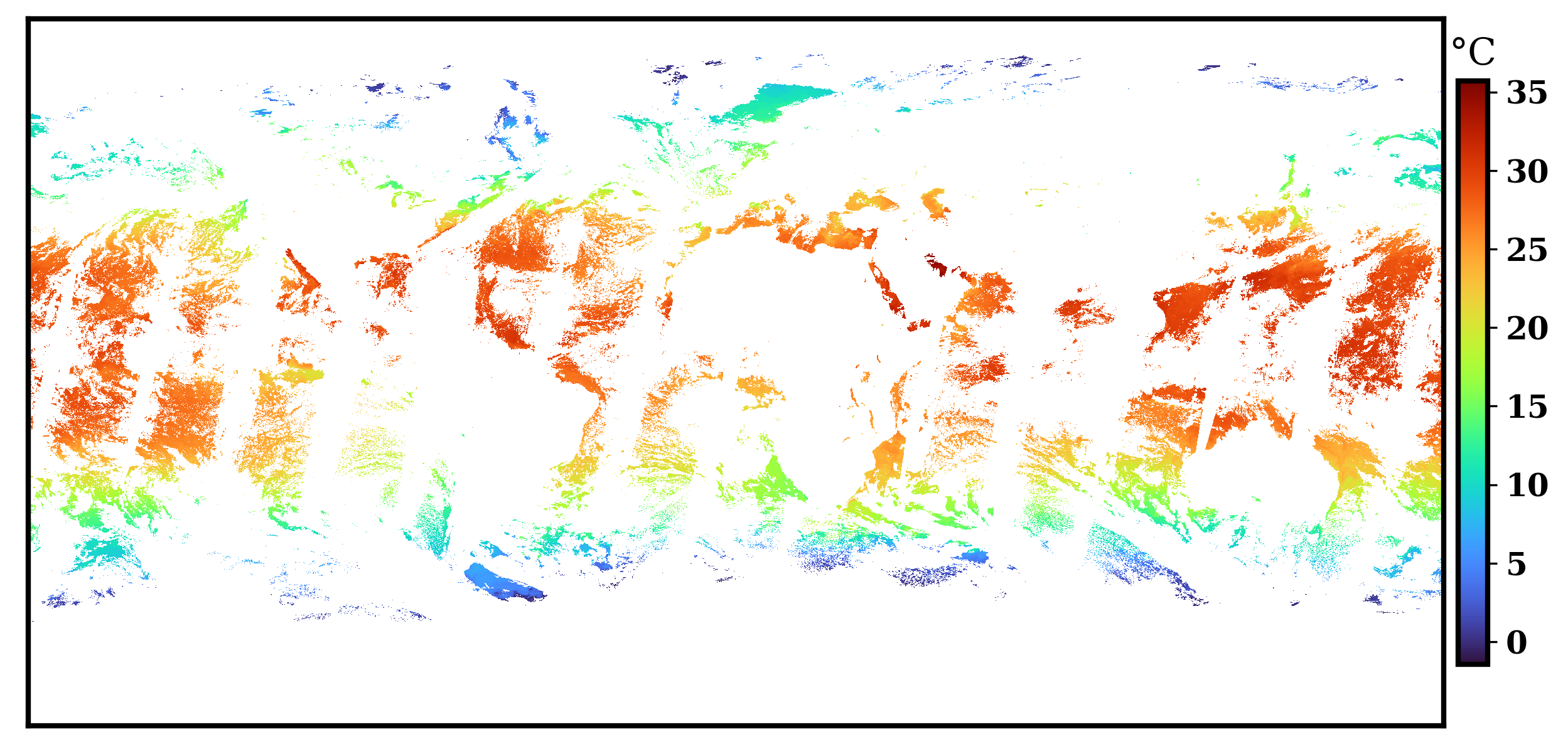}
    \caption{L3 daily MODIS file containing only high quality flagged pixels}
    \label{fig2_5}
\end{figure}

\begin{figure}[!ht]
	\centering
 	\includegraphics[width=1.0\linewidth]{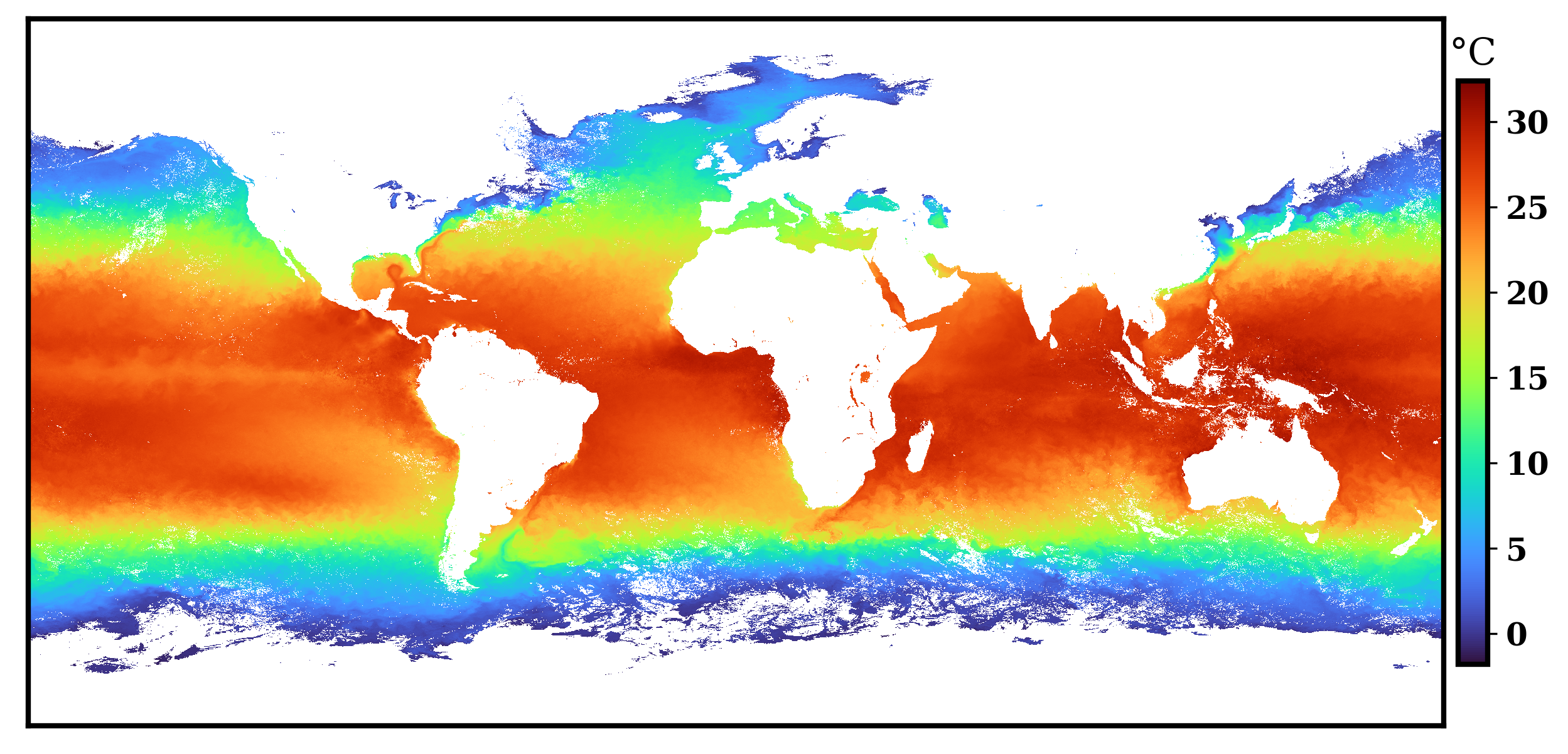}
    \caption{Monthly L3 MODIS image containing only high quality observations}
    \label{fig2_6}
\end{figure}

The MODIS Aqua Level 3 SST Thermal IR Monthly 4km V2019.0 product \cite{modis-day, modis-night} comes with latitude and longitude coordinates, SST values and per pixel quality measurements denoting when contamination is likely. The grid is equidistant rectangular, a match with the AMSR-E grid but at a finer original resolution. Of the over thirty million pixels for an entire day of 4 km MODIS pixels, 90\% of them in the random sample selected here are deemed contaminated and filtered out Figure \ref{fig2_5}. This contrasts with the 50\% loss of AMSR-E pixels. This great loss in pixels due to quality is attributed to cloud contamination. To compensate, we use the monthly product Figure \ref{fig2_6} where a greater amount of time has transpired, allowing for a higher probability of clean global coverage. A randomly sampled MODIS monthly image yields 50\% loss, in line with the AMSR-E daily product and much improved upon relative to the daily MODIS observation files.

\subsection{Ground Truth Measurements}
For a source of ground truth data, we selected the “In Situ Analysis System” (ISAS) dataset obtained from the University of California’s Argo repository and produced by a consortium of French institutions \cite{gaillard2016situ}. An important constraint for this work was to obtain only the surface level measurement of temperature at the highest frequency available during the years of both AMSR-E and MODIS. These products are provided in a gridded format are used to observe temperature measurements at many depth levels. In the publication attached to the ISAS dataset \cite{gaillard2016situ}, the target physical quantity is steric height and ocean heat content; with these as their target output, gridded depth-dependent temperature is stored as a byproduct. The 0.5 degree monthly dataset is presented in a Mercator projection, slightly different than the AMSR-E and MODIS grids. Mercator lines of longitude have a uniform distance in between them; the distance between latitudes from the equator changes. Identical to AMSR-E, we re-grid this data to the MODIS grid and coordinate system.

\subsection{Treatment}

This work is an extension of \cite{larson2023discerning}. The F2F code base provides an step-by-step approach to the fundamentals of the materials and methods applied within. As such, it is a key of part of the work and has been made openly accessible at \url{https://github.com/albertlarson/f2f} (accessed on 6 June 2023). The scripts follow the logical order of that paper. Concretely, this work builds upon that foundation. The notebooks found at \url{https://github.com/albertlarson/f2f_sst} follow the simple process of extracting the data, transforming the data, feed the data into a transformation system (dcrrnn), and then evaluating the performance of the system.

The treatments we apply to the data are several configurations of one common concept: neural networks. Neural networks are not new, but the growth of graphical processing units (hardware) has enabled them to flourish in software. Neural networks are a type of learned representation. A structure is fed connected input and target pairs. Based on the predictive quality of the initial network structure, an error between the neural network output and the target occurs. This error is in turn fed to an optimization algorithm that iteratively and slightly alters each “neuron” of the initial network structure until it reaches a designated optimal state. Via many small calculations and the simultaneous application of statistical mechanics, neural networks are known to provide qualities like that of a brain, such as capturing spatial eccentricities and temporal changes in sets of related images. Neural networks are applied to a range of tasks from the more mundane such as learning a quadratic equation, to the more cutting edge, like extreme event forecasting or cancer detection. 

Transfer learning has become commonplace in the field of machine learning \cite{weiss2016survey}. Transfer learning places an emphasis on creating reusable treatment structures for others to build on top of without inadvertently causing the audience to get lost in possibly unimportant details. We employ transfer learning to create a complex configuration with a relatively short learning curve. The neural network is characteristically deep, convolutional, residual, and regressive. Our construct is inspired by the work of residual networks \cite{he2016deep}. However, our problem is one of a regressive nature. Sea surface temperature has a continuous temperature range that it exists within. This is a notable difference to some of the more common introductory neural network examples, such as those associated with the MNIST and CIFAR datasets where the number of possible outputs is very small \cite{xiao2017fashion, krizhevsky2010convolutional}. Loss functions associated with regressive problems are constrained to just a couple: mean absolute error (MAE) and mean squared error (MSE). The calculation of the loss function must be differentiable. This is due to the optimization component of neural networks. The literature is rich with publications regarding neural network optimizers, as well as the general mechanics of neural nets \cite{andrychowicz2016learning, kingma2014adam}.

Once neural network architecture and hyperparameters are chosen, training and validation data is loaded into the network. While training the neural network, close observation is made of the reduction in error between training input and output as the neural network begins to optimize or learn. We also monitor the validation dataset at each training iteration. The learning process stops once the training and validation data has been passed through the network a certain number of times, or epochs. When prototyping or pilot-testing the experiment set to be carried out, one should test with a very short number of epochs and a larger sum of epochs to see where good performance meets fast time of computation.  

After training, the optimized neural network structure is intentionally frozen. Before the point of freezing, the neurons of the network can be adjusted for optimization, like a student asking a teacher for advice when studying. The frozen state and inference imposed upon it is like a student being prompted with a pop quiz and no teacher assistance. This test or input data are similar enough to the training that the teacher believes the student will have success in passing the test according to the selected merit (mean squared error, the loss function). After the test, the performance of the model is evaluated and a decision is made regarding next logical steps in the research.

A neural network can become biased to its training inputs. It starts to memorize the training dataset, which does not make for a generally applicable algorithm. Avoidance of biasing comes at the cost of variance \cite{geman1992neural}. Applying dropout is one technique to systematically prevent system bias by simply “turning off” a certain percentage of random neurons at each iteration of the algorithm \cite{hinton2012improving, srivastava2014dropout}. Another approach is the application of early stopping. The loss function of a neural network typically looks like a very steep curve down to a flat bottom. Rather than allow the network to persist in the flat bottom for long and become overfit, simple logic can be employed to stop training early when the network shows evidence that it has reached an optimal state. Percentage of data split between training and testing proportions is another relevant training hyperparameter. A larger proportion of the dataset being part of the training portion could lead to overfitting of the model and lack of generalized predictability. On the other hand, insufficient training data might lead to an inability to adequately characterize the reality of the data pairs.

\begin{figure}[!ht]
	\centering
 	\includegraphics[width=1.0\linewidth]{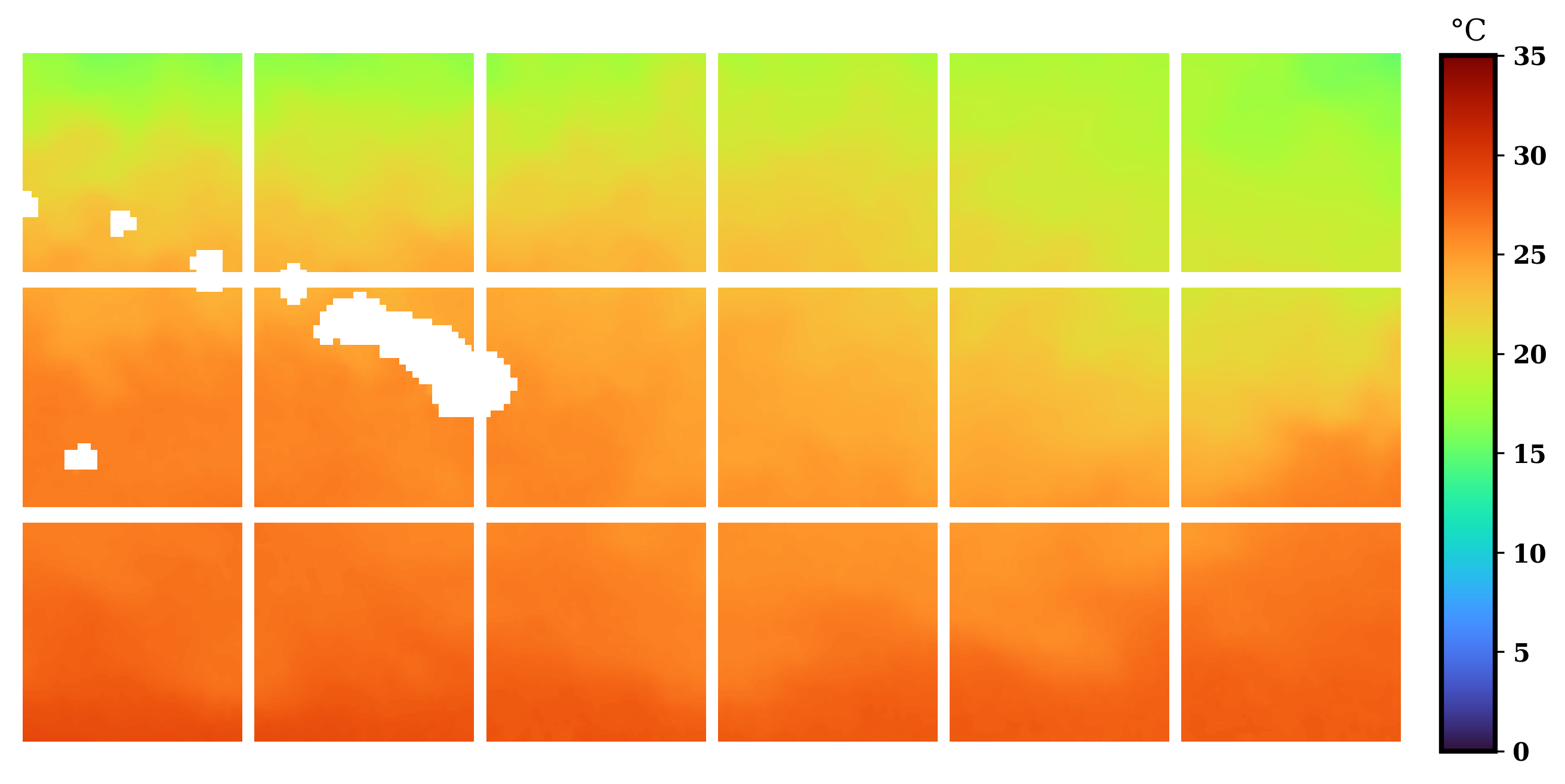}
    \caption{Sample training monthly observation; January 2010 MODIS day observation of the Hawaiian Islands; segmented into 100 x 100 pixel regions.}
    \label{fig2_7}
\end{figure}

The image sets subjected to treatment are on the large side computationally. Holding many one million or nine million pixel images within the memory of a single graphical processing unit becomes intolerable to the device. One could elect to use multiple GPUs or a compute node with a great provision of memory. Here, we constrain the experiment to a single GPU and cut the images up into smaller pieces of square data. Our patch size is fixed at 100 x 100, though this is a tunable hyperparameter. Figure \ref{fig2_7} shows a Pacific Ocean study region, highlighting Hawaii and regions east. While this image is too large to process directly in the neural network, we can solve this problem by creating the eighteen patches of 100 x 100 pixels, representing the 300 x 600 pixel region under observation. 

Neural networks do not function when nan values are present in any of the images. We enacted a broad treatment to the AMSR-E and MODIS images, computing the mean of the entire image, excluding the nan values. Then, where the nan values are present, we replace them with the mean value. This has the convenient byproduct of introducing into the neural network many training pairs where the input and output are simply comprised of the average global SST value as obtained via the AMSR-E and MODIS instrument.

\section{Results}
We randomly selected a single calendar year from the available time series where AMSR-E, MODIS, and ISAS overlap. Of this year's worth of data, we settled on nine different cases consisting of three different locations (Atlantic Ocean, Pacific Ocean, and Indian Ocean) and three different observation windows (day, night, mean averaged of day and night).
We train the neural network on the first ten months of the year, validate with the eleventh month, and test with the twelfth month. However, we did not intend for this system to be deterministic or biased in nature. Therefore, we shuffle the training pairs to confuse the network and promote regularization \cite{kumawat2022shuffleblock}. A training session runs for 100 epochs. Each image in the geographically constrained time series is 300 pixels x 600 pixels in size, divided up into eighteen 100 x 100 pixel segments to incrementally feed the neural net.

\begin{figure}[!ht]
	\centering
 	\includegraphics[width=1.0\linewidth]{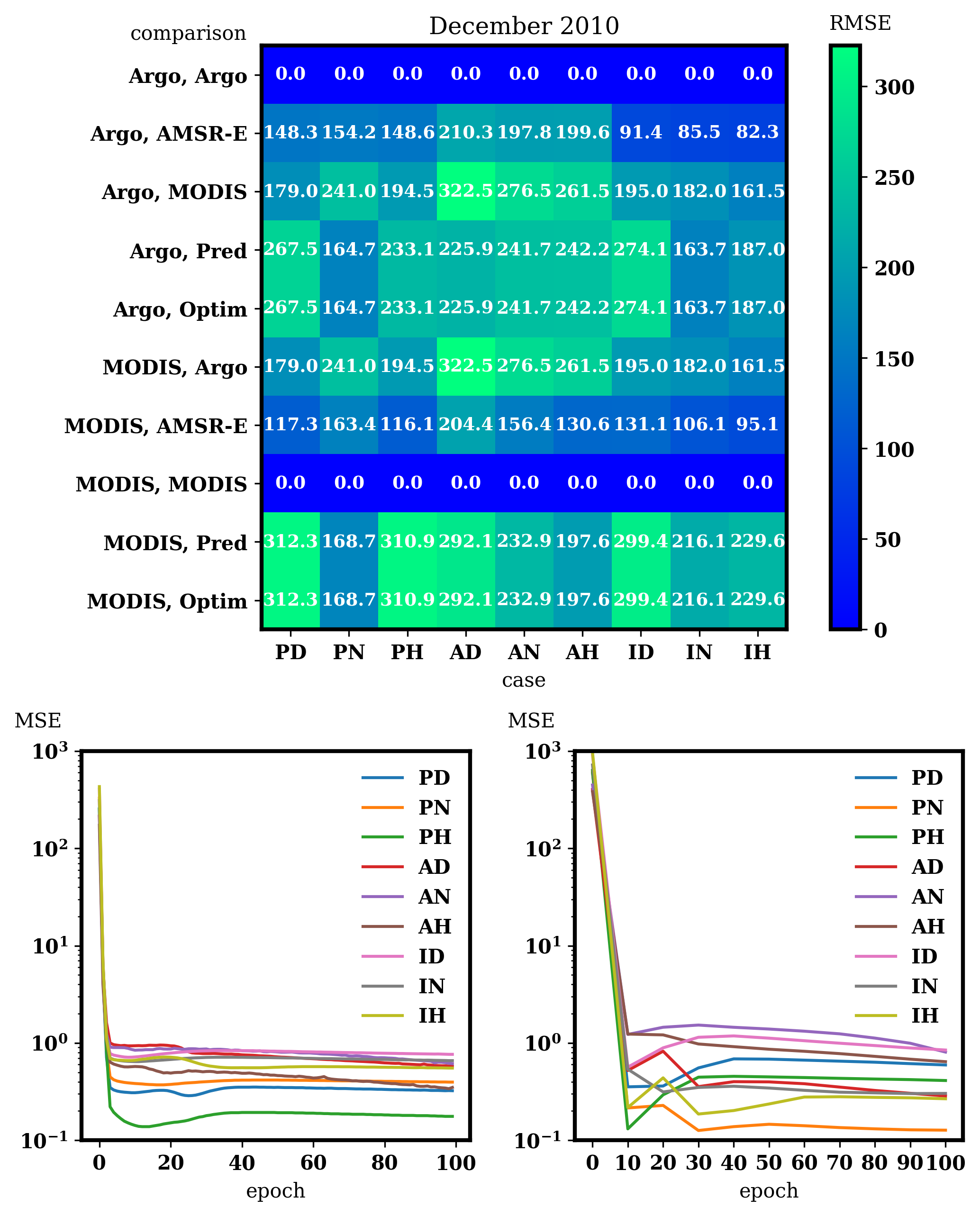}
    \caption{December 2010. The top panel shows the RMSE matrix for all datasets (input and outputs) for each of the nine cases. Two bottom panels are MSE plots of training and validation losses during neural network training. \textbf{A}tlantic, \textbf{P}acific, and \textbf{I}ndian Oceans segmented by monthly \textbf{D}ay, \textbf{N}ight, and \textbf{H}ybrid}
    \label{fig2_8}
\end{figure}

Results of the nine cases are illustrated in Figure \ref{fig2_8}. The upper panel is a test matrix. Each of the ten rows in the test matrix represent a tuple of datasets that are compared. All of the different data products we use are compared here to the ISAS data and the MODIS data as benchmarks. Columnwise, the test matrix is delineated into each of the nine cases (experiments) that we perform. For example, PD stands for Pacific Day, AN means Atlantic Night, and IH Indian hybrid. The hybrid is simply a mean of the day and night images for a given month. Take the PD column. We compute the RMSE between the complete 300 x 600 Pacific scene for each of the comparison tuples. In row 1, column 1, the RMSE value of the (Argo, Argo) comparison for the Pacific Day experiment comes out to zero. This is expected, because there should be no error between two identical datasets. Root mean squared error is not computed in locations where land is present.

The bottom two panels are graphs of the in-the-loop performance of the dcrrnn algorithm as the learning process occurs. The loss function, sometimes otherwise referred to as the cost function, for dcrrnn is mean squared error (MSE). This calculation is related to the RMSE metric used in the upper panel. Though the computations are different, the results are not perfectly related. The cause of this discrepancy is a function of a constraint for the training process. If an input or target image has any non-numerical pixels present, we have to replace them in some way. The cause of nans are largely attributed in this case to a location that is partially or completely land, or is a pixel contaminated by clouds or rainfall. We retain a global land mask for use before and after the training process to enable the removal of the temporarily-filled land pixels, which gives way to the clean RMSE test matrix. Our fill mechanism for each of the training and validation squares is to use the local mean of that square as the fill value. As such, there are always the same number of pixels factored into each 100 x 100 square during the neural network training process and in particular the calculation of in-the-loop training loss. 

Pred in the test matrix refers to the neural network's prediction of the unseen test data, the single month of data. It is fed into the trained network as eighteen 100 x 100 squares, but then recombined into one 180,000 pixel array. Land locations are added back on top. No other compute mechanism is employed. The Optim dataset has an extremely faint bandpass filter on top of the Pred dataset. Simply put, if there are any pixels in the Pred result from the neural network that is outside of the three standard deviations from the mean of the image, they are converted to nans and deemed erroneous. In this experiment, this filter has little effect. One would see more drastic effects should the trigger for filtering data from Pred to Optim change from three standard deviations to two, one, or less standard deviations from the mean. The risk of using a bandpass filter is that much of the interesting nuanced information can be filtered away. This feature was implemented during the experiment phase in response to the analyst's acknowledgement that artefact in the experiments were occurring along the coast.

\begin{figure}[!ht]
	\centering
 	\includegraphics[width=1.0\linewidth]{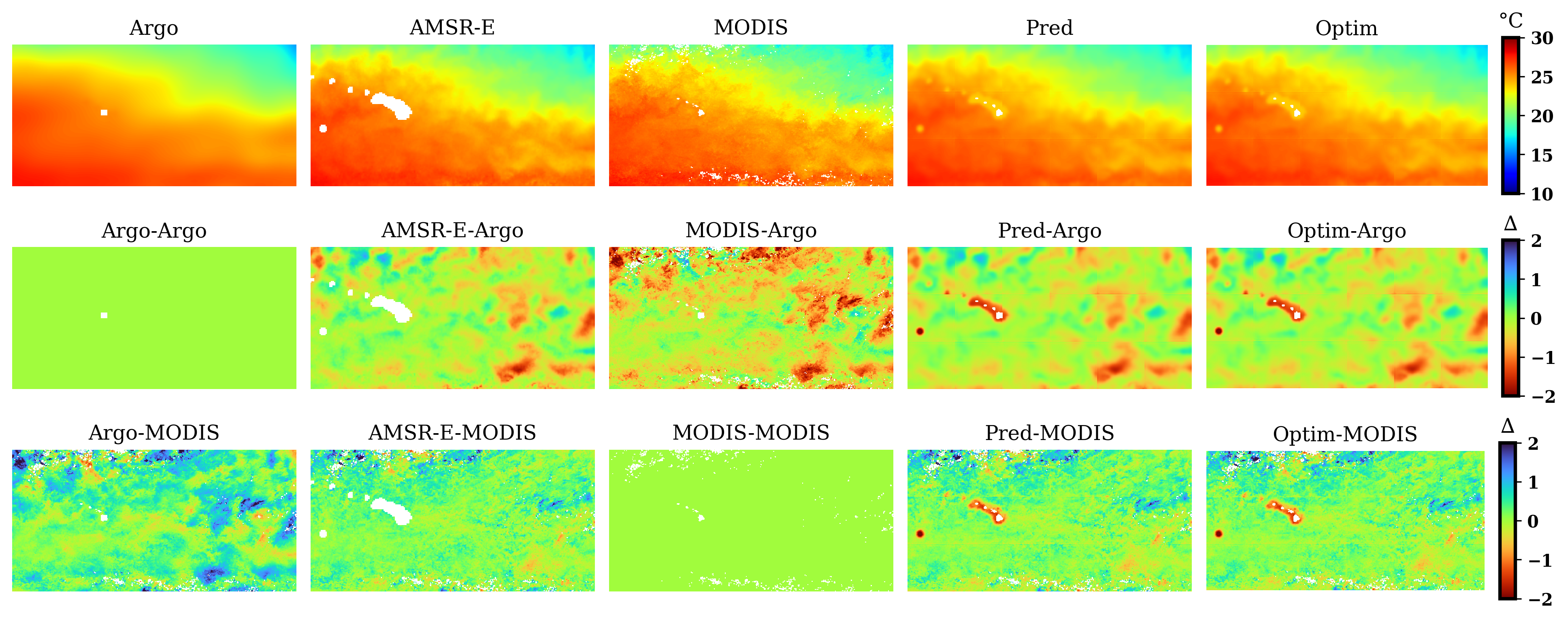}
    \caption{Relatively “good” perceptual change, Pacific Night case, axes are pixels}
    \label{fig2_9}
\end{figure}

\begin{figure}[!ht]
	\centering
 	\includegraphics[width=1.0\linewidth]{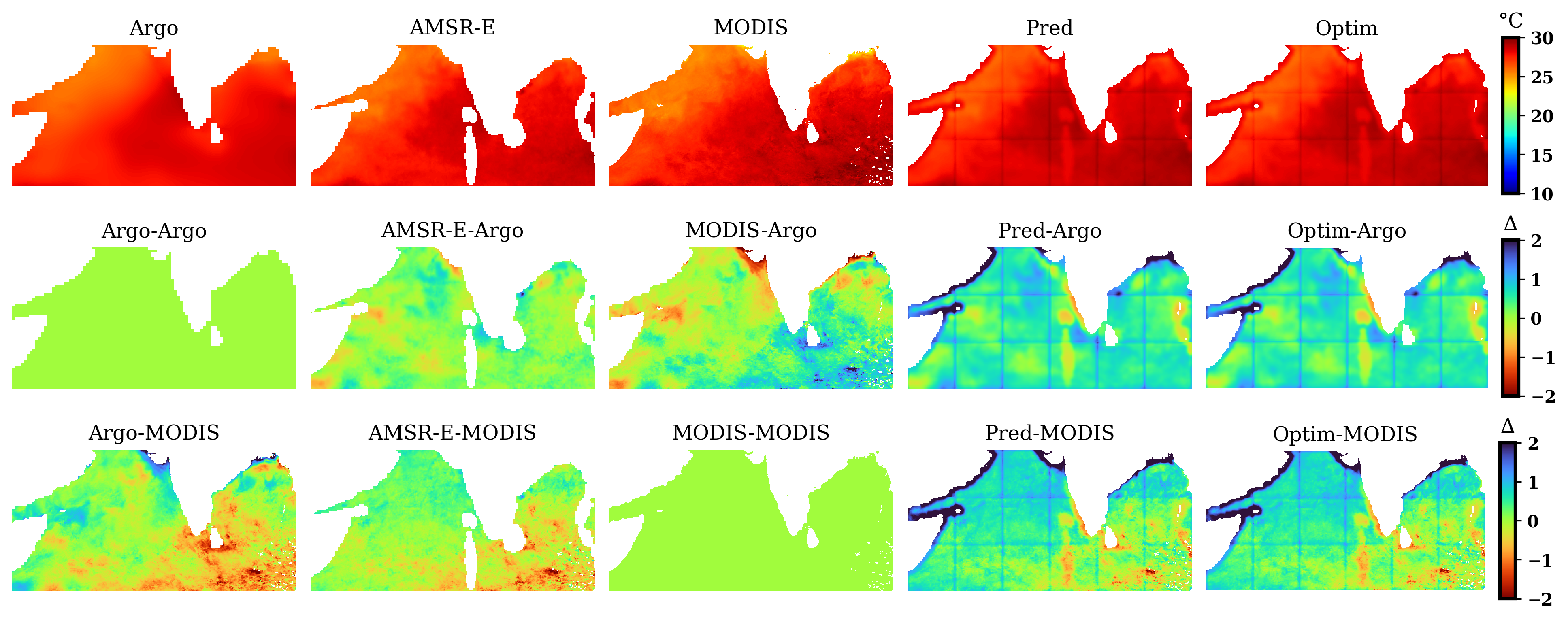}
    \caption{Relatively “poor” perceptual change, Indian Day case, axes are pixels}
    \label{fig2_10}
\end{figure}

\section{Discussion}

With every case summarized in Figure \ref{fig2_8}, the RMSE between Optim and MODIS is higher than the AMSR-E input. In occasional instances, the test case of December does make an Optim output that is closer to Argo than the input AMSR-E. In some cases, though, it makes a worse performing product with regards to Argo than either AMSR-E or MODIS. See Figures \ref{fig2_9} and \ref{fig2_10} for samples of how the RMSE translates to actual transformation of the images.  In the second row of the first column, one sees all green. This solid green color demarcates a clear relationship between the two compared datasets. In this case, that is because this square is the difference between Argo and itself. Considering all images in row 2 of Figure \ref{fig2_9}, the second column where AMSR-E is compared to Argo seems to be the most green, denoting a closeness. Pred and Optim are both covered in green; however, the bandpass filter in Optim is not being utilized. Therefore, we see the coastal artefact around the Hawaiian islands present as a result of dealing with nan values due to land. We are hesitant to use a liberal bandpass filter, because there's a real risk of deleting good dynamic data. This fact has lead us to an open question on the most appropriate way to handle water cycle data. It is evident that a gridded ocean product where land has no numerical representation isn't acceptable, and solutions like replacing values with local means isn't a true solution but a workaround. The most obvious answer is to create a unified global surface temperature dataset. However, this is less than best for the hydrology community. An adequate replacement specification for land surface temperature might be soil moisture or calculated surface and subsurface flows.

Figure \ref{fig2_10} represents a dcrrnn experiment conducted over the Arabian Sea and Bay of Bengal; it also illustrates a less desirable outcome than that of Figure \ref{fig2_9}. The telltale signal of underparameterization is evident in columns four and five of Figure \ref{fig2_10}, most notably in rows 2 and 3. As mentioned earlier, because of compute constraints, we can not load the entire image into the neural network at one time. We have to break it up into squares. Here there are clear horizontal and vertical boundary lines. Furthermore, there is the appearance of much colder estimates of SST, driven by artefact from the coastal regions. It appears that edges were systematically not able to be resolved based on the configuration of the dcrrnn structure and quantity of data fed into the system.

\begin{figure}[!ht]
	\centering
 	\includegraphics[width=1.0\linewidth]{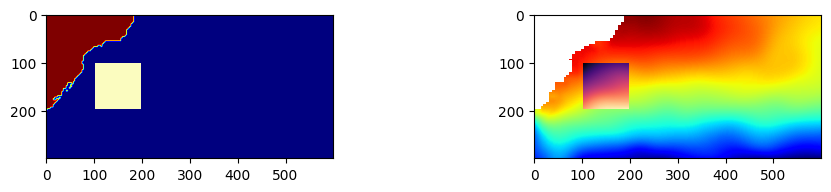}
    \caption{Atlantic Night, single patch selection overlain on top of the the land mask (left) and SST field (right)}
    \label{fig2_11}
\end{figure}

\begin{figure}[!ht]
	\centering
 	\includegraphics[width=0.5\linewidth]{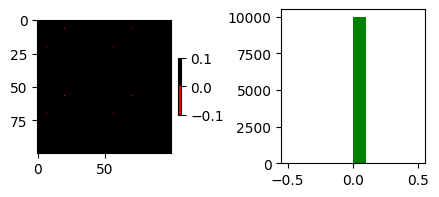}
    \caption{Atlantic Night, single 100 x 100 pixel patch real/nan Boolean (left) and histogram of real/nan (right)}
    \label{fig2_12}
\end{figure}

We determined that just looking at the nine cases in one constant way was not considerate enough of potential unseen indirect effects or confounding variables \cite{pearl2018book, pearl2022direct}. As an additional measure, we dug deeper into the night only observations of the Atlantic ocean. We selected a single 100 x 100 pixel that is completely over ocean within the Atlantic region. Our selection is seen in Figure \ref{fig2_11}. Figure \ref{fig2_12} illustrates that all pixels in the selection are real numbers (left). The presence of the color red would denote nan pixels present. This point is reinforced on the right panel of Figure \ref{fig2_12} with the single vertical line histogram denoting all real pixels.

\begin{figure}[!ht]
	\centering
 	\includegraphics[width=1.0\linewidth]{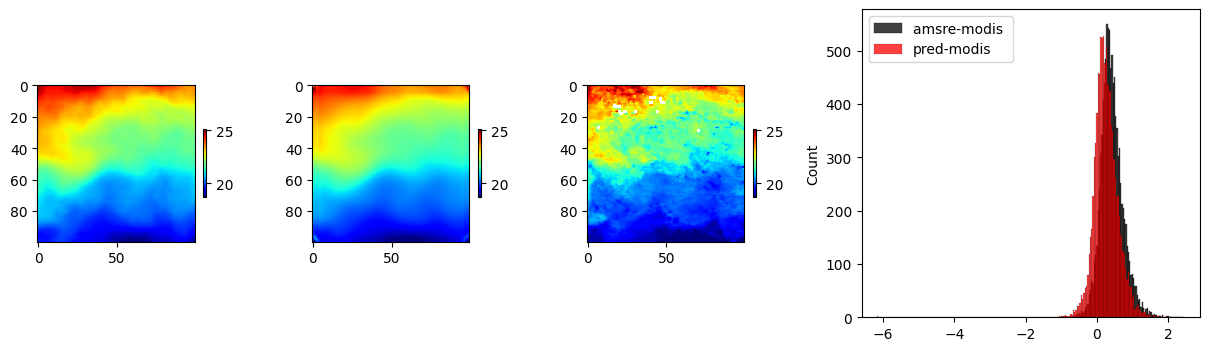}
    \caption{Atlantic Night, AMSR-E (col 1), Pred (col 2), MODIS (col 3), histogram of differences (col 4)}
    \label{fig2_13}
\end{figure}

We extend the number of epochs from 100 as compiled in Figure \ref{fig2_8} to 500 in this more-zoomed-in observation. A larger amount of training epochs, a smaller dataset size, and no nans present are strategic measures away from the baseline experiments to improve the outcome. Results of this experiment are illustrated in Figure \ref{fig2_13}. With this configuration, we find that the network reduces the RMSE difference between AMSR-E and MODIS by ~20\%, bringing the RMSE between AMSR-E and MODIS from 53.4 down to 44.0 between Pred and MODIS. However, there is no major perceptual change between the AMSR-E image and the Pred image. It is a promising result that there is an improvement in the results here relative to the nine larger cases. This performance indicates that rectification of the performance issues we are seeing can be further alleviated by shrinking the target size relative to the input, or by increasing the input size. As the output size shrinks, the problem converges with that of the earlier dcrrnn work \cite{larson2023discerning}.

Wherever land and coastal regions are present, the neural network alone struggles. This is due to the nature of the land sea boundary layer in all these datasets. At the presence of land, the raw data (as they are downloaded as .nc files) are given a non-number (nan) designation. For the purpose of training a neural network using the convolutional flavor, images with zero nans are needed. Otherwise the software fails catastrophically. As referenced earlier, steps were taken during the training process to circumvent the presence of land by substituting those pixels temporarily with the local mean value. Another option is the application of the substitution of the nan values with the mean as computed by the entire “scene” or day. There is a risk that the substitution of these values introduces a source of structured noise. This noise might be one factor leading to the higher than zero RMSE scores denoted in Figure \ref{fig2_8}. Furthermore, it is possible that this structured noise is hindering training of the neural network process itself. This is unfortunate, because coastal regions are the stage for a variety of interesting SST events such as boundary currents. At the same time, hydrologists and global health professionals have a growing stake in the influence of the ocean. Sea level rise and saltwater intrusion are two hallmark conditions present at the coastal interface \cite{mccarthy2018rapid}. Beyond sea surface temperature, ocean color and sea surface height are variables intimately linked to the coastal environment \cite{jutla2010tracking, panthi2022saltwater, piecuch2020likely}. We hypothesize that global datasets considering these parameters have the same challenges when conformation to a gridded structure and neural network process are prescribed. We believe that some harmony via integration of land and data sets would be a next discrete step to evaluate the impact on training and inferring with Flux to Flow. A good candidate is surface and subsurface flow as producted by the Global Land Data Assimilation System (GLDAS). GRACE, VIIRS, and SMAP all have land products available as well as oceanic products. It's possible that if the data is ocean focused, perhaps leaving the land observations in their raw state with no physical scaling would be the best practice.

Our images are single channel inputs, and can be considered grayscale pictures. They are slightly different though; grayscale pictures are usually digitized as pixels with numbers between 0 and 1. When displaying SST images, measurements of physical properties, we use a colormap with minimum and maximum based on what we know to be the physical limits of the parameter itself. Nevertheless, it is closely related to a grayscale image. In grayscale computer vision tasks such as this set of experiments, the use of mean squared error as a loss function has  audience of skeptics \cite{wang2009mean}. There are examples where the loss function optimized in a neural network drops in value significantly but sees no improvement in the quality of the image. Alternate loss functions to the standards built into PyTorch are available \cite{kastryulin2022image}. Without alteration, these functions require the inputs to be either between 0 and 1 (grayscale) or 0 and 255 (color images). Another avenue is the pursuit of physics-based loss functions \cite{psaros2022meta}. Some data engineering is definitely needed in future iterations. As we have experience with the surface and subsurface flow rasters and these are land viewing, we think the combination of SST and GLDAS as predictors of streamflow in coastal regions is a suitable discrete next step. This next step will force us to decrease the number of target outputs, as there is a paucity of point measurements in streamflow monitoring relative to that of the interpolated target ISAS data used herein. 

Only a fraction of the available data was observed in this study. The ISAS Argo dataset comes as a single compressed file; we extracted  the surface layer only. There is great value in consideration of SST depth layers; however, it was outside the scope of our investigation. Furthermore, we studied monthly time series images of all three raw datasets. AMSR-E and MODIS each have near complete global pictures within two to three days. These datasets are then transformed in different ways and can lose fidelity by various types of decimation such as re-gridding from swaths to squares, uncertainty in formulas used for conversion from base input to high level physical parameter, or by forms of compression. 

We took a naive approach to the problem. It is common to initialize a model with a historical known bias, and make slight inferences based upon the long-term mean. This tightly constrains the problem to the known past environment. We did not do that. We actively attempt to root out and prevent any sort of deterministic bias, and see under a tough conditioning what the dcrrnn algorithm does. This leads to some less than desirable results; however, our zoom in shows that when the complexity of the system is relaxed, the algorithm improves according to the RMSE target. In a future study, it would be helpful to start the training with a long-term bias of SST for an entire year based on, for example, the Operational Sea Surface Temperature and Ice Analysis time series dataset produced by the Group for High Resolution Sea Surface Temperature \cite{ghrsstostia}.

\section{Conclusions}

Sea surface temperature (SST) is an essential climate variable. A better understanding of SST equates to a better understanding of global hydrology and the interactions of water as it moves around the hydrosphere in liquid, solid, and vapor forms. The advent of satellite SST observation has allowed for the study of large scale phenomena otherwise invisible. The beginning of the 21st century marked a new frontier in the measurement of SST via the Aqua mission and Argo program. Recently, neural networks have changed the way that scientists consider modeling of the environment. In this study, we continued to develop Flux to Flow, an extract, transform, load, treat, and evaluation framework based around a deep convolutional residual regressive neural network (dcrrnn). We extended its additional functionality of streamflow prediction to transform one Aqua instrument dataset into another: AMSR-E observations towards MODIS observations. 

We focused on three large oceanic regions: Indian, Pacific, and Atlantic. With each of the three locations comprised of eighteen 100 x 100 pixel image pairs per month, and ten months of training data, the neural network struggles to transform AMSR-E into MODIS. When we relax the amount of data fed into dcrrnn, looking at a single 100 x 100 pixel image pair per month and ten months of training data, the network is statistically better able to transform AMSR-E into MODIS data. Provided these results, we believe that a next discrete step is to focus on coastal areas where the hydrology and oceanography are closely linked. We would like to investigate the performance of dcrrnn in predicting the streamflow of a river when it does and does not consider oceanic behavior in the adjacent ocean. We hypothesize that merging ocean and land datasets will not only ease the challenges associated with handling non-numbers, but that the streamflow prediction will benefit from the unique signatures present in the ocean data alongside the land measurements of the water cycle.

\bibliographystyle{unsrt}  
\bibliography{references}

\end{document}